\documentclass[aps,prx,twocolumn,groupedaddress]{revtex4-2}

\usepackage{amsmath,amssymb,graphicx,dcolumn,bm,multirow, upgreek, hyperref}

\begin{document}

\title{Postselection-free controlled generation of a high-dimensional orbital-angular-momentum entangled state}


\author{Suman Karan}
\email[]{ karans@iitk.ac.in}
\affiliation{Department of Physics, Indian Institute of Technology Kanpur, Kanpur, UP 208016, India}

\author{Radhika Prasad}
\affiliation{Department of Physics, Indian Institute of Technology Kanpur, Kanpur, UP 208016, India}

\author{Anand K. Jha}
\email[]{akjha@iitk.ac.in}
\affiliation{Department of Physics, Indian Institute of Technology Kanpur, Kanpur, UP 208016, India}

\date{\today}

\begin{abstract}

High-dimensional entangled states in orbital angular momentum (OAM) basis offer several unique advantages for  quantum information applications. However, for the optimal performance of a given application, one requires a generation technique for OAM entangled states that is completely postselection-free and fully controllable. Nonetheless, despite several efforts in the past, no such technique currently exists. In this article, we propose just such a technique and experimentally demonstrate postselection-free generation of up to about 150-dimensional OAM entangled states. We report the generation accuracy, which is a measure of the control, to be more than 98$\%$ for states with Gaussian and triangular OAM Schmidt spectra and up to 90$\%$ for the maximally-entangled OAM states, which have rectangular spectra.

\end{abstract}

\maketitle

\section{Introduction}
A Laguerre-Gaussian (LG) mode $LG^{l}_{p}(\rho,\phi)$ is characterized by indices $l$ and $p$, referred to as the orbital angular momentum (OAM) mode index and the radial mode index, respectively. A photon in an LG mode with index $l$ carries OAM of $l\hbar$, where $l$ is an integer \cite{allen1992pra}. Thus the OAM of a photon provides a discrete infinite-dimensional basis \cite{yao2011advoptics} as opposed to the two-dimensional polarization basis \cite{forbes2019avsquantum}. The use of high-dimensional entangled states provides higher error tolerance for quantum-key distribution \cite{bechmannpasquinucci2000prl, nikolopoulos2006pra}, higher security against  eavesdropper attacks in quantum cryptographic schemes \cite{cerf2002prl}, increased information capacity \cite{wang2012natphotonics, bozinovic2013science}, enhanced robustness of communication protocols in noisy environment \cite{ecker2019prx, zhu2021avsquantum}, and stronger violations of generalized Bell's inequalities \cite{leach2009optexp, dada2011natphys}. High-dimensional entangled states also have important implications for dense coding \cite{wang2017optica, hu2018scienceadvance}, quantum teleportation \cite{luo2019prl, hu2020prl}, entanglement swapping \cite{zhang2017naturecomm, takeda2015prl}, and supersensitive measurements \cite{jha2011pra1, chen2014lsa, asban2019pnas}.

In the past, there have been several efforts at producing high-dimensional OAM-entangled  states using spontaneous parametric down-conversion (SPDC), in which a pump photon of higher frequency splits into two photons of lower frequencies, called the signal and idler. It is known that for the optimal performance of a quantum information protocol \cite{bouchard2018quantum, ding2017npjqinformation}, one needs to generate OAM entangled states in a completely postselection-free and controlled manner. However, all the past efforts have only yielded generation techniques that are either postselection-free but not fully controllable \cite{pires2010prl, kulkarni2018pra, torres2003pra, kulkarni2017natcomm} or fully controllable but not postselection-free \cite{leach2009optexp, dada2011natphys, wang2017optica, hu2018scienceadvance, luo2019prl, wang2015nature, hu2020prl, zhang2017naturecomm, takeda2015prl, jha2011pra1, vaziri2003prl}.  More recently, there have been experimental \cite{liu2018pra, liu2020pra, kovlakov2018pra} and theoretical \cite{bornman2021aqt, xu2022optexp, miatto2012epjd} efforts at developing generation techniques that are fully controllable but postselection-free only with respect to $p=0$ mode detection. These techniques involve single-mode-fiber based OAM detection system, which only detects photons with radial mode index $p=0$ \cite{qassim2014josab, agrawal_fiber_optics}. However, it is known that the proportion of $p\neq 0$ radial modes in the OAM entangled state generated by SPDC is typically much larger than the $p=0$ mode \cite{miatto2011pra, zhang2014pra}. Hence, while these techniques \cite{liu2018pra, liu2020pra, kovlakov2018pra, bornman2021aqt, xu2022optexp, miatto2012epjd} can have limited use in some applications, they are not postselection-free. Thus, to the best of our knowledge, there is no existing technique for a truly postselection-free generation of high-dimensional OAM entangled states with full control. In this article, we experimentally demonstrate generation of truly postselection-free OAM entangled states with control.

\section{Theory}

The two-photon state generated by type-I SPDC in the transverse momentum basis can be written as \cite{karan2020jopt, torres2003pra, walborn2010physreports}
 \begin{align}\label{two_photon_wavefunc_momen}
 | \psi_{\rm tp} \rangle = \iint V({\bf q}_s,{\bf q}_i)\Phi({\bf q}_s,{\bf q}_i)|{\bf q}_s\rangle_s |{\bf q}_i\rangle_i d^2 {\bf q}_s d^2 {\bf q}_i,
 \end{align}
where $p$, $s$ and $i$ denote pump, signal and idler, respectively. $V({\bf q}_s,{\bf q}_i)$ represents the pump field amplitude, $\Phi({\bf q}_s,{\bf q}_i)$ represents the phase matching function, and  $|{\bf q}_s\rangle_s$ and $ |{\bf q}_i\rangle_i $ denote the state vectors in the transverse momentum basis of the signal and idler, respectively. In the LG basis, we can write  $| \psi_{\rm tp} \rangle$ as
 \begin{align}\label{two_photon_wavefunc_oam}
 \qquad |\psi_{\rm tp}\rangle = \sum_{l_s,p_s}\sum_{l_i,p_i}C^{l_s, p_s}_{l_i,p_i} |l_s, p_s\rangle_s |l_i, p_i\rangle_i,
 \end{align}
where $l_p$, $l_s$ and $l_i$ are the OAM mode indices of the pump, signal, and idler, respectively \cite{mair2001nature}. $|l_s, p_s\rangle_s$ is the state of the signal photon with indices $l_s$ and $p_s$, etc. Thus, using Eqs.~(\ref{two_photon_wavefunc_momen}) and (\ref{two_photon_wavefunc_oam}), we write $C^{l_s, p_s}_{l_i,p_i}$ as \cite{miatto2011pra, yao2011njp}:
\begin{multline}\label{C_ls_li}
 C^{l_s, p_s}_{l_i,p_i} =  A\int\!\!\!\!\int d^2 {\bm q}_s  d^2 {\bm q}_i  V({\bm q}_s, {\bm q}_i)\Phi({\bm q}_s,{\bm q}_i) \\
 \times  \left[ LG^{l_s}_{p_s}({\bm q}_s)\right]^{*}\left[ LG^{l_i}_{p_i}({\bm q}_i)\right]^{*},
\end{multline}
where  $ LG^{l}_{p}({\bm q})= \langle {\bm q}| l , p \rangle  $ is the momentum basis representation of the LG mode \cite{torres2003pra}. Now, using the cylindrical polar coordinates, ${\bm q}_s =(q_{sx}, q_{sy})=(\rho_s {\rm cos} \phi_s , \rho_s {\rm sin} \phi_s)$, ${\bm q}_i =(q_{ix},q_{iy})=(\rho_i {\rm cos} \phi_i , \rho_i {\rm sin} \phi_i)$, $d^2 {\bm q}_s = \rho_s d \rho_s d \phi_s$, and $d^2 {\bm q}_i = \rho_i d \rho_i d \phi_i$, we write $C^{l_s, p_s}_{l_i,p_i}$ as  
\begin{align}
 C^{l_s, p_s}_{l_i,p_i}&= \int\!\!\!\!\int_{0}^{\infty}\!\!\!\!\int\!\!\!\!\int_{-\pi}^{\pi} \rho_s\rho_i d\rho_s d \rho_id \phi_s d \phi_i  V(\rho_s, \rho_i,\phi_s,\phi_i) \notag \\ \times \Phi(& \rho_s, \rho_i,\phi_s,\phi_i) \left[ LG^{l_s}_{p_s}(\rho_s,\phi_s)\right]^{*} \left[ LG^{l_i}_{p_i}(\rho_i,\phi_i)\right]^{*}.  \label{c_ls_li_polar}
\end{align}
$ P^{l_s}_{l_i}$ is the probability of detecting the signal and idler photons with OAM $l_s \hbar$ and $l_i \hbar$, and it is given by
\begin{align}
 P^{l_s}_{l_i} = \sum^{\infty}_{p_s=0}\sum^{\infty}_{p_i=0} |C^{l_s,p_s}_{l_i,p_i}|^2. \label{pls_li}
\end{align}   
When $l_p =0$, conservation of OAM leads to $l_s = -l_i$ \cite{mair2001nature}, and in this case, Eq.~(\ref{two_photon_wavefunc_oam}) takes the following Schmidt decomposed form:
\begin{align}
|\psi\rangle_{\rm tp}= \sum_l \sqrt{S_l} |l\rangle_s |-l\rangle_i. \label{schmidt-decomposed form}
\end{align}
Here $S_l= P^{l}_{-l}$ is called the OAM Schmidt spectrum. Using Eqs.~(\ref{c_ls_li_polar}) and (\ref{pls_li}), and employing the identity $ \sum^{\infty}_{p=0}(LG)^{l}_{p}(\rho)(LG)^{*l}_{p}(\rho')= \frac{1}{\pi}\delta(\rho^2 - \rho'^2)$ \cite{jha2011pra2}, we write $S_l$ as (for a more detailed derivation, see \cite{kulkarni2018pra}):
\begin{multline}\label{sl}
S_l =P^{l}_{-l}= \frac{1}{4 \pi^2} \int\!\!\!\!\int_{0}^{\infty}\rho_s \rho_i \Bigg|\int\!\!\!\!\int_{-\pi}^{\pi} V(\rho_s, \rho_i,\phi_s,\phi_i)  \\ 
\times\Phi(\rho_s, \rho_i,\phi_s,\phi_i) e^{-i l(\phi_s - \phi_i)} d \phi_s  d \phi_i \Bigg|^2  d\rho_s  d \rho_i .
\end{multline}
%
%
%
%
%
From Eq.~(\ref{sl}), we find that $S_l$ is determined by the pump field amplitude $V(\rho_s, \rho_i,\phi_s,\phi_i)$ and the phase-matching function $\Phi(\rho_s, \rho_i,\phi_s,\phi_i)$. Therefore, by selecting an appropriate $V(\rho_s, \rho_i,\phi_s,\phi_i)$ and $\Phi(\rho_s, \rho_i,\phi_s,\phi_i)$, one can generate any target OAM Schmidt spectrum. However, solving Eq.~(\ref{sl}) for finding the analytical expressions for $V(\rho_s, \rho_i,\phi_s,\phi_i)$ and $\Phi(\rho_s, \rho_i,\phi_s,\phi_i)$ for a target OAM Schmidt spectrum $S_l$ is mathematically a very difficult problem. So, our technique for finding the required $V(\rho_s, \rho_i,\phi_s,\phi_i)$ and $\Phi(\rho_s, \rho_i,\phi_s,\phi_i)$ for a target spectrum is based on optimization and feedback.

We note that $S_l$ expressed in Eq.~(\ref{sl}) includes contributions from all possible radial modes since the probability $P^{l_s}_{l_i}$ in Eq.~(\ref{pls_li}) has been defined as a sum over all the radial modes of the signal and idler photons. Thus $S_l$ is the true OAM Schmidt spectrum. This is in contrast with several recent techniques \cite{liu2018pra, liu2020pra, kovlakov2018pra, bornman2021aqt, xu2022optexp, miatto2012epjd}, in which the spectrum is taken as the probability of only the $p=0$ mode. We find that the spectral content observed with a $p=0$ mode detector is only a small fraction of the true spectrum (see Appendix~\ref{supple: postselection} for a detailed study regarding this.) Thus our generation technique, which is based on measuring the $S_l$ in Eq.~(\ref{sl}) is truly postselection-free.

\begin{figure}[t!]
\centering
\includegraphics[scale=0.73]{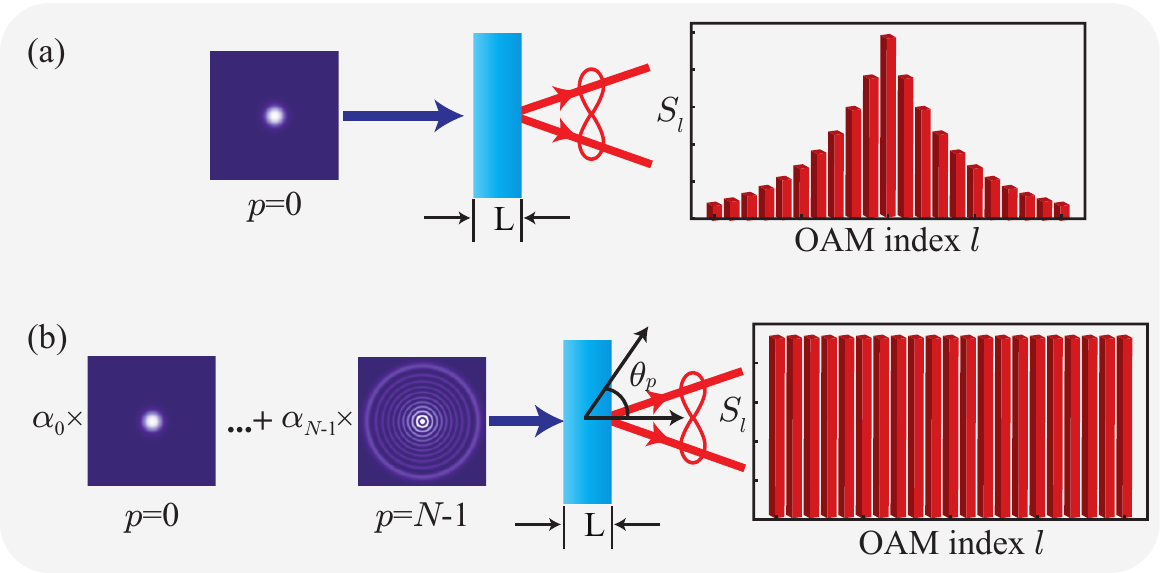}
\caption{ (a) Generation of the OAM entangled state using a Gaussian pump field amplitude $(LG^{l_p=0}_{p=0})$ and collinear phase matching. (b) Postselection-free controlled generation of OAM entangled states with noncollinear phase matching and a pump field in the form of a superposition of modes with different radial indices.}
\label{conceptual_diagram}
\end{figure}

Figure~\ref{conceptual_diagram} illustrates our technique. A Gaussian pump field with the collinear phase matching provides a very limited control over generating the OAM Schmidt spectrum. However, as shown in Figure~\ref{conceptual_diagram}(b), for a controlled postselection-free generation, we use non-collinear phase matching and take the pump field amplitude $V(\rho_s, \rho_i,\phi_s,\phi_i)$ in the form of a coherent superposition of LG modes with different $p$-indices:
\begin{figure*}[t!]
\centering
\includegraphics[scale=0.90]{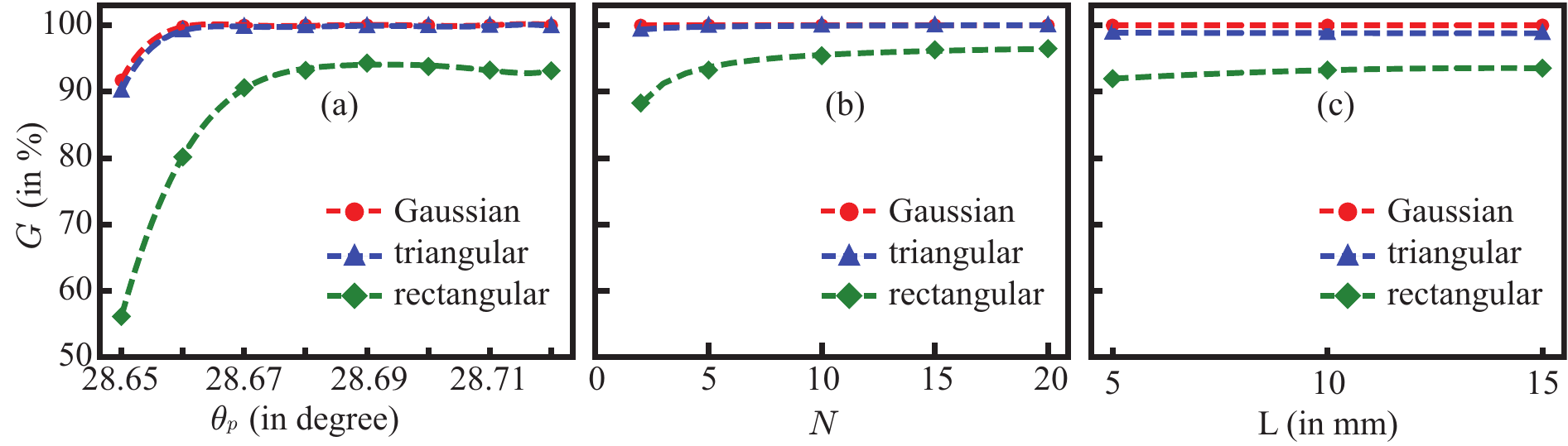}
\caption{Numerical plots of the dependence of generation accuracy $G$ for three different OAM Schmidt spectra: Gaussian, rectangular, and triangular. (a) Plot of $G$ as a function of the phase matching angle $\theta_p$ for $N=5$ and $L=10$ mm. (b) Plot of $G$ as a function of the maximum number of radial modes $N$ for  $L=10$ and $\theta_p= 28.71^{\circ}$. (c) Plot of $G$ as a function of the crystal thickness $L$ for $N=5$ and $\theta_p= 28.71^{\circ}$. Dashed lines denote the best fit.}
\label{theory_thetap_mode_L_plot}
\end{figure*}
\begin{align}\label{pump_superposition}
V(\rho_s, \rho_i,\phi_s,\phi_i)=\sum^{N-1}_{p=0} \alpha_{p} LG^{l_p=0}_p(\rho_s, \rho_i,\phi_s,\phi_i),
\end{align}
where $\alpha_p$ is the complex coefficient, $N$ is the total number of radial modes in the superposition, and  
\begin{multline}
LG^{l_p=0}_p(\rho_s, \rho_i,\phi_s,\phi_i) = \left[\frac{w_p^2}{2 \pi}\right]^{1/2} L^0_p\left[ \frac{w_p^2 \rho_p^2}{2}\right]  \\
\times  \exp\left[-\frac{w_p^2 \rho_p^2}{4}\right] \exp\left[ i \pi p\right].
\end{multline}
Here $ L^0_p $ is the associated Laguerre polynomial, $\rho_p ^2 = \rho_s ^2 + \rho_i ^2 + 2\rho_s \rho_i \cos(\phi_s - \phi_i)$, and $w_p$ is the beam waist of the pump field. The phase matching function $\Phi(\rho_s, \rho_i,\phi_s,\phi_i)$ of Eq.~(\ref{sl}) can be written as \cite{karan2020jopt, kulkarni2018pra, walborn2010physreports}
\begin{align}
\Phi(\rho_s, \rho_i, \phi_s,\phi_i) = \mbox{sinc}\left(\frac{\Delta k_z L}{2}\right)\exp\left[i\frac{\Delta k_z L}{2}\right],
\end{align}
where $L$ is the thickness of the crystal, and $\Delta k_z$ is called the phase mismatch parameter. $\Delta k_z$ can be tuned by changing the phase matching angle $\theta_p$, which is the angle between the propagation direction of the pump field and the optic axis of the nonlinear crystal. For the detailed expression of $\Delta k_z$, see Appendix~\ref{supple: phase_matching_calculation}.

In our technique, we first numerically simulate a given OAM spectrum by optimizing the complex coefficients $\alpha_p$. For a $2D+1$ dimensional state, we define the coefficient of determination $R^2$ (see Ref.~\cite{glantz2001primer} Chap $6$) as
\begin{align}\label{cost}
R^2 =\left[1-  \dfrac{\sum_{l=-D}^{+D} \left(S_l^{\rm t} - S_l^{\rm o}\right)^2}{\sum_{l=-D}^{+D} \left(S_l^{\rm t}- \langle S_l^{\rm t} \rangle \right)^2 }\right]\times 100 \%,  
\end{align}
where $S_l^{\rm t}$ and $S_l^{\rm o}$  are the target and experimentally observed OAM Schmidt spectra, and $\langle S_l^{\rm t} \rangle$ is the mean of $S_l^{\rm t}$. For a target OAM Schmidt spectrum $S_l^{\rm t}$, we maximize $R^2$ by numerically optimizing $\alpha_p$ using a particle swarm optimization technique \cite{bonyadi2017evcomputation} implemented with the Python package `pyswarm'. We define the maximum value of $R^2$
\begin{align}
{\rm max}\{R^2\} \equiv G \label{generation accuracy}
\end{align}
as the generation accuracy $G$ of our technique. $G$ quantifies the degree of control that our technique has for generating a target OAM Schmidt spectrum. Finally, we use the numerically optimized coefficients $\alpha_p$ as starting points for further optimizing them experimentally through a feedback mechanism.
\begin{figure*}[!t]
\centering
\includegraphics[scale=0.93]{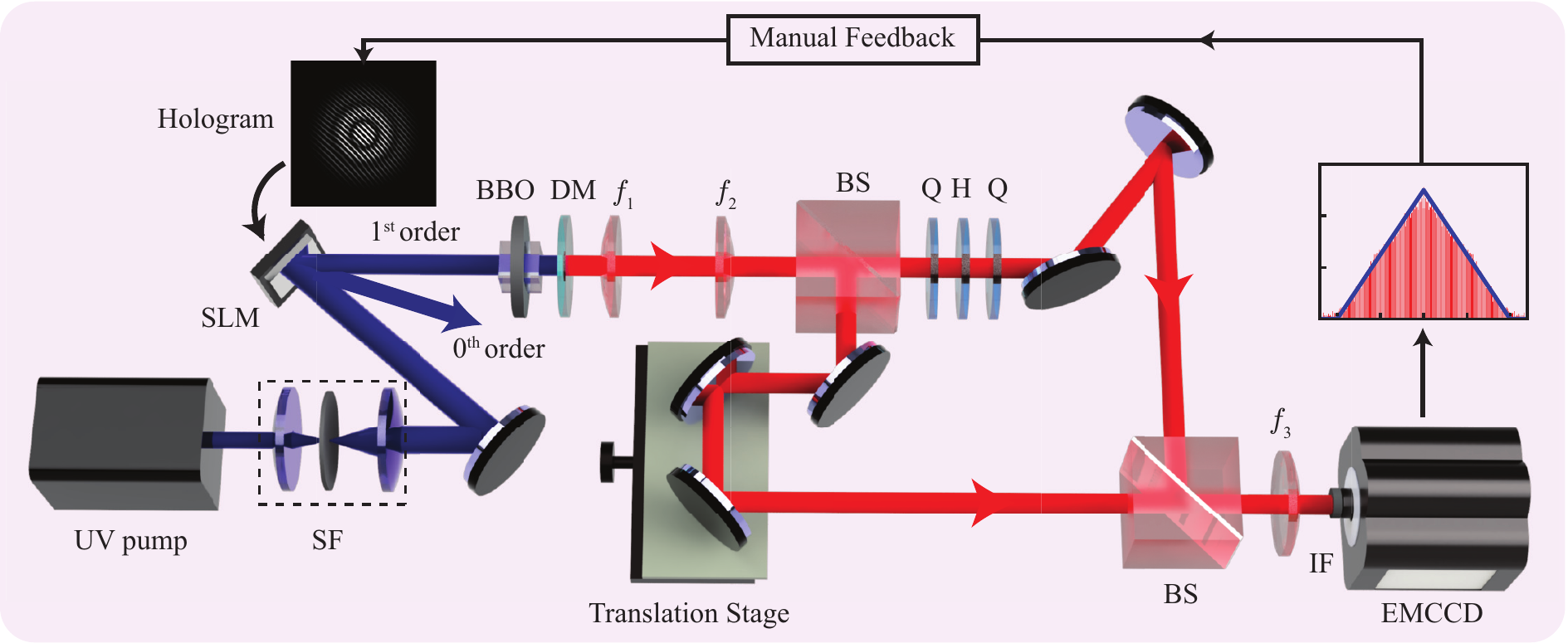}
\caption{Schematic of the experimental setup. SF: spatial filter, SLM: spatial light modulator, BBO: $\beta$- barium borate crystal, DM: dichroic mirror, BS: 50:50 non-polarization beam splitter, Q: quarter-wave plate, H: half-wave plate, IF: interference filter of central wavelength $810$ and $10$ nm frequency bandwidth. }
\label{exp_setup}
\end{figure*}

Figure~\ref{theory_thetap_mode_L_plot} presents the results of our numerical studies on how the generation accuracy $G$ depends on the phase matching angle $\theta_p$, the number of radial modes $N$, and crystal thickness $L$, for three different spectra, namely, Gaussian, triangular and rectangular. The rectangular spectrum is particularly interesting in that it represents a maximally entangled state. Nonetheless, it has been shown that for several quantum information applications non-maximally entangled states are preferred over the maximally entangled states \citep{bandyopadhyay2012pra, acin2012prl, alimuddin2023arxiv}. Therefore, we include the Gaussian and triangular spectra in our analysis. Figure~\ref{theory_thetap_mode_L_plot}(a) shows the plot of $G$ versus $\theta_p$. For these plots, we have taken $N=5$ and $L=10$ mm. We find that with the collinear phase matching condition, characterized by $\theta_p= 28.65^{\circ}$, it is not possible to achieve very high values of $G$ but at $\theta_p$ set for sufficiently large noncollinear down-conversion, $G$ can get close to 100$\%$ for the Gaussian and triangular spectra. However, for the rectangular spectrum, $G$ stays much lower. The main reason for this is that one requires a larger number of radial modes $N$ in the superposition in Eq.~(\ref{pump_superposition}) for producing spectrum such as rectangular which has sharp edges. Figure~\ref{theory_thetap_mode_L_plot}(b) shows the numerically simulated plot of $G$ as a function of $N$ for $L=10$ mm and $\theta_p= 28.71^{\circ}$. We find that as $N$ increases, $G$ increases. Furthermore, for Gaussian and triangular spectra, $N=5$ is enough to get $G$ close to 100$\%$ whereas for the  rectangular spectrum one requires at least $N=10$ to reach $G>95\%$. Finally, Fig~\ref{theory_thetap_mode_L_plot}(c) shows the plot of $G$ as a function of $L$ for $N=5$ and $\theta_p= 28.71^{\circ}$. We find that for a target spectrum, once $\theta_p$ and $N$ have been optimized, $G$ does not get much affected by $L$. 


\section{Experiments}

The experimental setup is depicted in Fig.~\ref{exp_setup}. We use a 100-mW Toptica TopMode  ultraviolet (UV) continuous wave (CW) laser of wavelength $\lambda_p= 405$ nm that is spatially filtered and incident on a  Holoeye Pluto-2-UV-099 spatial light modulator (SLM). The SLM is used for generating the pump field $V(\rho_s, \rho_i,\phi_s,\phi_i)$ of Eq.~(\ref{pump_superposition}) in the form of a superposition of $N$ radial modes using the method described by Arrizon et al. \cite{arrizon2007josaa}. For $N=5$, we use modes from $p=0$ to $ p=4$ with pump beam waist $w_p= 320~ \mu\rm{m}$. The pump field is made incident on a $\beta$-barium borate (BBO) crystal of transverse dimensions 10 mm $\times$ 10 mm and thickness $L=10$ mm. The crystal is kept on a goniometer to change the phase matching angle $\theta_p$. A Dichroic mirror (DM) is positioned just after the crystal to block UV while allowing the down-converted photon pairs to pass through. These photon pairs are incident on a Mach-Zehnder type interferometer \cite{kulkarni2017natcomm}, and the interferogram output is captured using an Andor iXon Ultra-897 electron-multiplied charged coupled device (EMCCD) camera with 512 $\times$ 512 pixel grids having pixel pitch of $ 16 \times 16 ~\mu {\rm m}^2$. Lenses with focal lengths $f_1 = 50~ {\rm mm}$, and $f_2= 200~ {\rm mm}$ are placed in a $4f$ configuration to image the crystal plane with a magnification of four. The Fourier transform of this image plane is obtained at the camera plane using another lens of focal length $f_3= 300$ mm. For capturing the interferograms, the camera acquisition time is kept at $20$ seconds. An interference filter (IF) of central wavelength 810 nm and bandwidth 10 nm is placed in front of the camera. For a target spectrum $S^{t}_l$, we use the optimized complex coefficients $\alpha_p$ from our numerical simulations as the starting point and produce the pump field $V(\rho_s, \rho_i,\phi_s,\phi_i)$ using the SLM. We then experimentally observe the generated spectrum $S^{o}_l$ with EMCCD using the two-shot technique described in Ref.~\cite{kulkarni2017natcomm} and  calculate $R^2$ using Eq.~(\ref{cost}). We then employ manual feedback in order to obtain an optimized set of $\alpha_p$ that maximizes $R^2$. For the optimization, we take $\alpha_p$ one-by-one, optimize its real and imaginary parts first in steps of $\pm 0.1$ and then in smaller steps. The experimentally observed value of $R^2$ after optimization is then taken as the generation accuracy $G$.

\begin{figure*}[!t]
\centering
\includegraphics[scale=0.94]{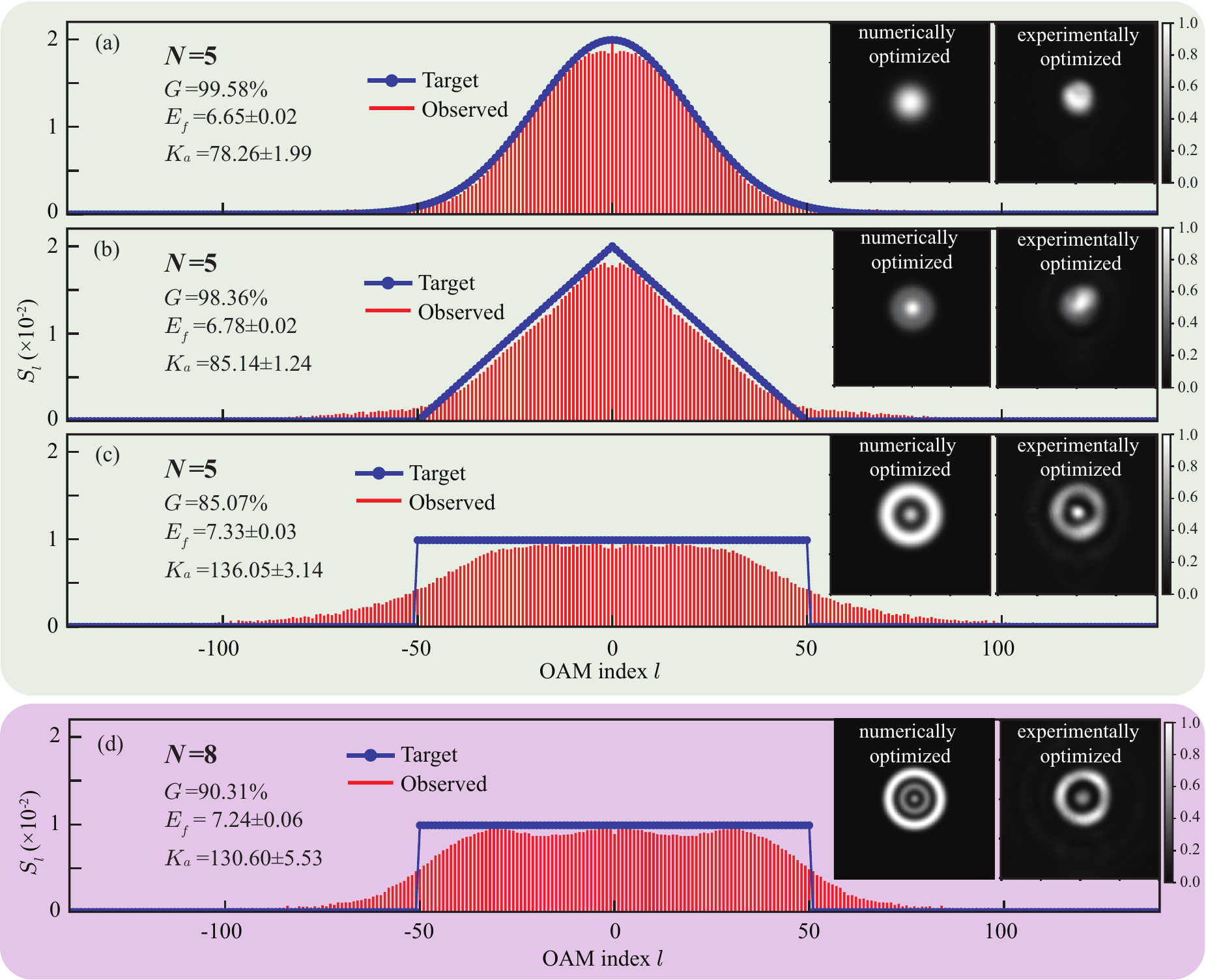}
\caption{ Experimentally generated OAM Schmidt spectra. (a), (b) and (c) Plots of target and experimentally observed OAM Schmidt spectra having Gaussian, triangular and rectangular shapes with $L=10$ mm, $\theta_p = 28.71^\circ$, and $N=5$. (d) Plot of target and experimentally observed rectangular OAM Schmidt spectrum with $L=10$ mm, $\theta_p = 28.71^\circ$, and $N=8$. The pump field intensities after the initial numerical and then the final experimental optimizations have been shown as insets.}
\label{L10mm_spectrumandpump}
\end{figure*}

We note that the strength of the pump field hitting the BBO crystal is only about 5 mW. We find that at this field strength, the relative probability of generating a four-photon state versus a two-photon state is less than $10^{-7}$ \cite{schneeloch2019jopt}. Therefore, the contribution due to four-photon effects in our experiment is negligible. We further note that although our technique is postselection-free in the OAM basis, the two-photon generation in time is only probabilistic due to the pump field being CW.

Figure~\ref{L10mm_spectrumandpump} presents our experimental results. Figures~\ref{L10mm_spectrumandpump}(a), \ref{L10mm_spectrumandpump}(b) and \ref{L10mm_spectrumandpump}(c) show the experimentally generated Gaussian, triangular and rectangular OAM Schmidt spectra with $N=5$. The pump field intensities for the three spectra after numerical optimization and then final experimental optimization have been shown as insets and the  superposition coefficients $\alpha_p$ after the final experimental optimization  have been reported in Appendix~\ref{supple:l10mm_exp}. The target standard deviation of the Gaussian spectrum is 20 while the target full width of the triangular and rectangular spectra is 100.

In order to verify that the states produced above are entangled, we use entanglement of formation $E_f$ as a certifier \citep{bennett1996pra}. An $E_f >0$ confirms that the state is entangled. We note that when the two-photon state is mixed, certifying entanglement through $E_f$ requires measuring correlations in at least two conjugate bases \cite{hill1997prl, wootters1998prl}. More recently, it has been shown that even the quantification of entanglement of high-dimensional mixed states can be done through only two measurements \cite{li2023prl}. However, for the case of pure two-photon states, entanglement can be verified by measuring correlation in only one basis \cite{bennett1996pra, wootters2001qcomputation} or of only one of the photons \cite{bhattacharjee2022njp, walborn2006nature}. In our technique, we use a spatially completely coherent pump field, which ensures that the generated two-photon state is pure \cite{jha2010pra, kulkarni2017josab, kulkarni2016pra}. Therefore, we calculate $E_f$ using the formula $E_f = - \sum_l S_l {\rm{log}}_2 S_l$ for the three spectra and find it to be $6.65 \pm 0.02$, $6.78 \pm 0.02$, and $7.33 \pm 0.03$.

The Schmidt number $K_a$ quantifies the dimensionality of generated sates and is calculated using the formula $K_a = 1/\sum_l S_l^2$ \cite{jha2011pra2}. For the three different target spectra reported in Figs.~\ref{L10mm_spectrumandpump}(a), \ref{L10mm_spectrumandpump}(b) and \ref{L10mm_spectrumandpump}(c), the corresponding Schmidt numbers are 78.26$\pm$1.99, 85.14$\pm$1.24 and 136.05$\pm$3.14, respectively. The generation accuracy $G$ for the three spectra is 99.58$\%$, 98.36$\%$, and 85.07$\%$, respectively. From the results presented in Figs.~\ref{L10mm_spectrumandpump}(a) - \ref{L10mm_spectrumandpump}(c), we find that there is always a difference between the numerically and experimentally optimized coefficients. This is because the numerically optimized results are for the ideal conditions, but in a real experiment, the conditions are usually quite different from being ideal. We also note that for the Gaussian and triangular spectra, $G$ is more than $98\%$ while for the rectangular spectrum, $G$ is only about $85\%$. As shown in Fig~\ref{theory_thetap_mode_L_plot} and also pointed out in the Theory section above, the main reason for the lower $G$ value for the rectangular spectrum is that a much larger $N$ is required for producing spectrum such as rectangular that has sharp edges. Although $N=5$ is sufficient for producing Gaussian and triangular spectra with almost perfect accuracy, one needs larger $N$ for producing rectangular spectrum.

Therefore, for the rectangular spectrum, we next perform experiments with $N=8$ radial modes, ranging from $p=0$ to $ p=7$. The other experimental details remain the same as mentioned above. Figure~\ref{L10mm_spectrumandpump}(d) shows the experimentally generated spectrum. The optimized coefficients are reported in Appendix~\ref{supple:l10mm_exp}. The experimentally measured values of $G$, $E_f$, and $K_a$ are  90.31$\%$, 7.24$\pm$0.06, and 130.60$\pm$5.53, respectively. We note that with $N=8$ modes, $G$ for the rectangular spectrum becomes more than $90\%$. A further increase in $N$ can increase $G$ even further. However, the maximum $N$ that can be implemented gets limited by two main factors. One is that as we increase the radial mode index $p$, the transverse size of the pump beam increases. As a result, for a given transverse size of the nonlinear crystal, we can only use a finite number of radial modes, and this puts an upper limit on $N$. Although one can increase $N$ by reducing the beam waist and thereby fitting more radial modes, a reduced pump beam waist results in a larger spread of the pump beam \cite{jha2010pra} making it difficult to capture the entire down-converted field by the detection system. The other factor is the algorithm for optimizing $\alpha_p$. In our experiment, the optimization of $\alpha_p$ involves a manual feedback. Therefore, increasing $N$ implies increasing the number of iterations in the feedback process and thereby increasing the experimental complexity and overall noise. Nevertheless, a feedback mechanism with automated controls can make the optimization much more efficient.  Therefore, with a nonlinear crystal with larger transverse size and an automated feedback mechanism, a much higher generation accuracy, even for a sharp-edged spectrum, should be easily achievable with our technique.

Finally, in order to highlight the control provided by our technique, we perform experiments with three different $L$, namely $L=5$ mm, $L=10$ mm, and $L=15$ mm, for producing Gaussian, triangular and rectangular spectra with two different spectral widths. (see Appendix~\ref{supple:l5mm_exp}, \ref{supple:l10mm_exp} and \ref{supple:l15mm_exp} for details). We find that different crystal thickness requires different phase matching conditions. Therefore, by optimizing $\theta_p$ and $\alpha_p$, one can generate a given target OAM spectrum for a wide ranges of crystal thickness. The only implication is that the optimization complexity becomes different at different crystal thickness. In Appendix~\ref{supple:l5mm_exp}, \ref{supple:l10mm_exp}, and \ref{supple:l15mm_exp}, we report generation of states with target dimensionality up to 200. By increasing $\theta_p$, the dimensionality of the generated state can be further increased, which ultimately gets limited by the collection aperture of the detection system.

\section{Conclusion}
In conclusion, we have proposed and demonstrated an experimental technique based on SPDC for postselection-free controlled generation of up to about 150-dimensional OAM entangled states. Ours is the first truly postselection-free technique for generating OAM-entangled states with full control. We note that the main limitation of our technique is that it  works only for the OAM-entangled states in the Schmidt-decomposed form. This is due to the fact that currently there is no efficient detector for measuring a general OAM-entangled state. With the advent of such a detector, our technique could in principle be extended to general OAM-entangled states. Furthermore, it has been shown that for several quantum information applications non-maximally entangled states are preferred over the maximally entangled states \citep{bandyopadhyay2012pra, acin2012prl, alimuddin2023arxiv}. Thus, our work can have important implications for high-dimensional quantum information.



\begin{acknowledgments}
We thank Girish Kulkarni and Manik Banik for helpful discussions. We acknowledge financial support from the Science and Engineering Research Board through grants STR/2021/000035 and CRG/2022/003070 and from the Department of Science $\&$ Technology, Government of India through Grant DST/ICPS/QuST/Theme-1/2019). SK thanks the University Grant Commission (UGC), Government of India for financial support.
\end{acknowledgments}



\begin{figure*}[]
\centering
\includegraphics[scale=1.18]{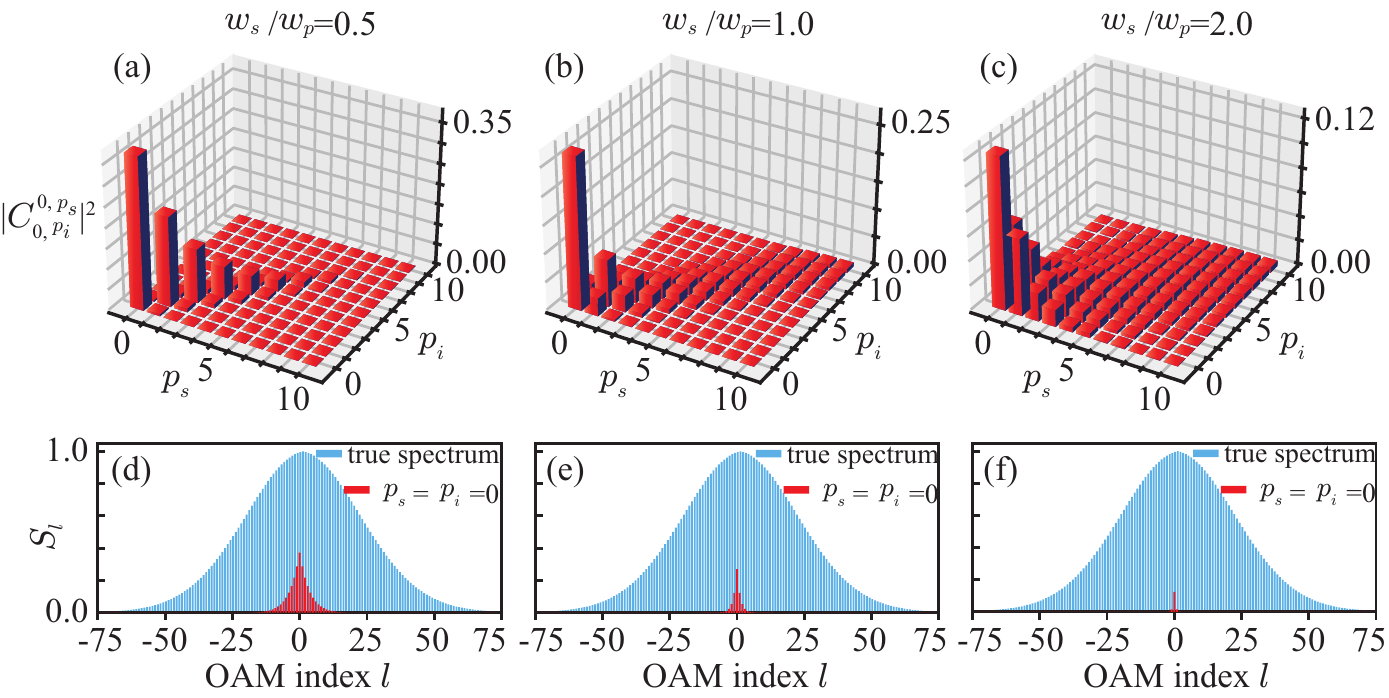}
\caption{The extent of postselection in detection techniques based on selecting only the $p=0$ mode. (a),(b),(c) are plots of $C^{0, p_s}_{0,p_i}$ for $w_s/w_p =0.5$, $1$, and $2$, respectively. (d), (e), (f) are the plots of the true Schmidt spectrum for $w_s/w_p$=0.5, 1, and 2, respectively. For the above plots, we have taken L=$10$mm, $\theta_p= 28.71^{\circ}$ and $w_p= 320 \mu$m. }
\label{fig: p_distribution_oam_spectrum}
\end{figure*}

\appendix

\section{extent of postselection in techniques based on detecting $p=0$ radial mode}\label{supple: postselection}

$|C^{l_s,p_s}_{l_i,p_i}|^2$ is the probability of detecting the signal and idler photons in LG modes with indices ($l_s, p_s$) and ($l_i, p_i$), respectively. The expression for $C^{l_s, p_s}_{l_i,p_i}$  is given in Eq.~(\ref{c_ls_li_polar}), where  $LG^{l_S}_{p_s}(\rho_s,\phi_s)$ is the projection mode for the signal photon. We write $LG^{l_S}_{p_s}(\rho_s,\phi_s)$ as 
\begin{align}
&LG^{l_S}_{p_s} (\rho_s,\phi_s)= \left[\dfrac{w_s~ p_s!}{2 \pi \left(|l_s|+ p_s\right)!}\right]^{\frac{1}{2}} \left[\dfrac{w_s \rho_s}{\sqrt{2}}\right]^{|l_s|} L^{|l_s|}_{p_s}\left[\dfrac{w_s ^2 \rho_s ^2}{2}\right] \notag \\  & \quad \times {\rm exp}\left[-\dfrac{w_s ^2 \rho_s ^2}{4}\right] {\rm exp}\left[il_s \phi_s + i \pi \left(p_s - \dfrac{|l_s|}{2}\right)\right],
\end{align}
where $L^{|l_s|}_{p_s}[\cdots]$ is called the  associated Laguerre polynomial, and $w_s$ is beam waist of the projected mode basis in which the signal photon is being detected. $LG^{l_i}_{p_i}(\rho_i,\phi_i)$ in Eq.~(\ref{c_ls_li_polar}) has a similar expression.  We take $w_i = w_s$ and calculate $|C^{l,p_s}_{-l,p_i}|^2$ using Eq.~(\ref{c_ls_li_polar}).

Figures~\ref{fig: p_distribution_oam_spectrum}(a), \ref{fig: p_distribution_oam_spectrum}(b) and \ref{fig: p_distribution_oam_spectrum}(c) show the numerically calculated plot of $|C^{0,p_s}_{0,p_i}|^2$ as a function of $p_s$ and $p_i$ for $w_s/w_p = 0.5$, $1$, and $2$, respectively. We take  $L=10$ mm, $\theta_p = 28.71^{\circ}$, $w_p = 320 \mu$m. We normalize  $|C^{0,p_s}_{0,p_i}|^2$ within the space $p_s =0 $ to $p_s=10$ and $p_i=0$ to $p_i=10$,  such that the total probability is equal to one. Next, Figs.~\ref{fig: p_distribution_oam_spectrum}(d), \ref{fig: p_distribution_oam_spectrum}(e) and \ref{fig: p_distribution_oam_spectrum}(f) show the plots of  OAM Schmidt spectra $S_l$ as a function of $l$ calculated using two different methods. One is the true Schmidt spectrum $S_l$, which contains contribution due to all the radial modes. It is derived in Eq.~(\ref{sl}) and is given  by: $S_l= P^{l}_{-l} = \sum^{\infty}_{p_s=0}\sum^{\infty}_{p_i=0} |C^{l,p_s}_{-l,p_i}|^2$. The other one is the probability $|C^{l,p_s=0}_{-l,p_i=0}|^2$ which contains contribution only due to $p_s=0$ and $p_i=0$ modes.  We find that the OAM Schmidt spectrum observed with a detection scheme that is sensitive only to $p=0$ modes \cite{liu2018pra, liu2020pra, kovlakov2018pra, bornman2021aqt, xu2022optexp, miatto2012epjd} entails strong postselection, and that the extent of this postselection increases with increasing $w_s/w_p$.

\section{Calculation of the phase mismatch parameter}\label{supple: phase_matching_calculation}
The phase matching function in Eq.~(10) of the manuscript is defined as $
\Phi(\rho_s, \rho_i, \phi_s,\phi_i) = \mbox{sinc}\left(\frac{\Delta k_z L}{2}\right)\exp\left[i\frac{\Delta k_z L}{2}\right]$.  Here $\Delta k_z$ is called the phase mismatch parameter and is given by \cite{karan2020jopt}
\begin{align}\label{deltakz}
\Delta k_z = k_{sz} + k_{iz} - k_{pz},
\end{align}
where
\begin{align*}
k_{pz} &= -\alpha_p q_{px} + \eta_p K_{p0} - \frac{ \left[ \beta_p^2 q^2_{px} + \gamma_p^2 q^2_{py}\right]}{2 \eta_p K_{p0}},\\
k_{sz} &=  n_{so} K_{s0} - \frac{1}{2 n_{so} K_{s0}}(q_{sx}^2+q_{sy}^2), \\
k_{iz} &=  n_{io} K_{i0} - \frac{1}{2 n_{io} K_{i0}}(q_{ix}^2+q_{iy}^2),
\end{align*}
Here $k_{sz}$ represents the $z$-component of the wave vector for signal field, etc. $K_{j0}= \frac{2\pi} {\lambda_{j}}$ with  $j=s,i$ and $p$, and  $n_{jo}$ and $n_{je}$ are the refractive indices for the ordinary and extraordinary polarizations. For type-I SPDC, the polarization of the pump photon is extraordinary while that of the signal and idler photons is ordinary. The quantities $\alpha_p$ , $\beta_p$, $\gamma_p$ and $\eta_p$  are given by 
\begin{align*}
\alpha_p = \frac{(n^2_{po}- n^2_{pe})\sin\theta_p \cos\theta_p}{n^2_{po}\sin^2\theta_p + n^2_{pe} \cos^2 \theta_p},\\
\beta_p = \frac{n_{po} n_{pe}}{n^2_{po}\sin^2\theta_p + n^2_{pe} \cos^2 \theta_p},\\
\gamma_p = \frac{n_{po} }{\sqrt[]{n^2_{po}\sin^2\theta_p + n^2_{pe} \cos^2 \theta_p}},\\
\eta_p = \frac{n_{po} n_{pe} }{\sqrt[]{n^2_{op}\sin^2\theta_p + n^2_{pe} \cos^2 \theta_p}}, 
\end{align*} 
In our experiment, we use BBO crystal for down-conversion, and the values of $n_{po}$, $n_{pe}$ for BBO can be obtained from the dispersion relation reported in Ref. \cite{eimerl1987jap}. For degenerate SPDC, we can write $K_{s0}\approx K_{i0} \approx K_{p0}/{2}$ and $n_{so} = n_{io}$. Numerically, we find  that for our experimental parameters, $\alpha_p \approx 0$ and $\beta_p \approx \gamma_p \approx 1$.  Thus,  $\Delta k_z$ can be written in the transverse momentum basis as 
\begin{align}
\Delta k_z &= K_{p0} \left[ n_{so} - \eta_p \right]  \notag \\  &- \frac{1}{2 \eta_p K_{p0}} \left[ \rho_s ^2 + \rho_i ^2 - 2\rho_s \rho_i ~{\bm cos}\left(\phi_s - \phi_i\right)\right].
\end{align}

\begin{figure*}[t!]
\centering
\includegraphics[scale=0.86]{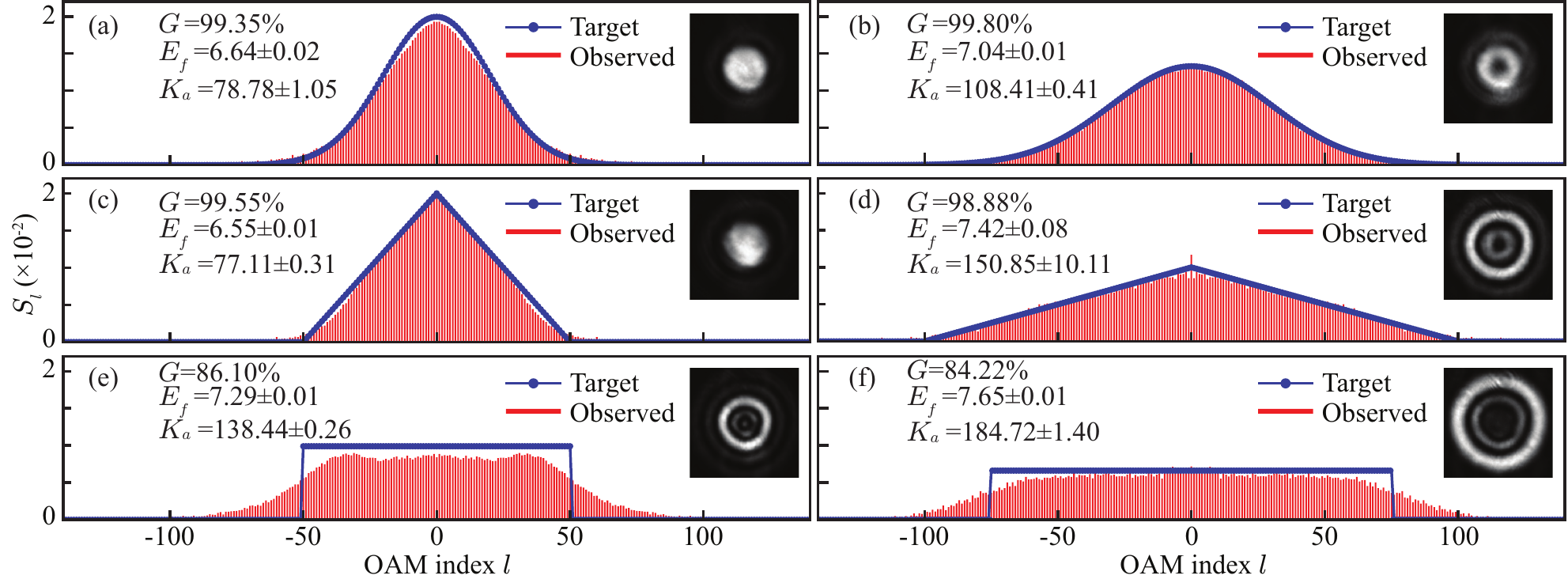}
\caption{Target and experimentally observed OAM Schmidt spectra with $L =5$ mm, and $\theta_p= 28.69^{\circ}$ with two separate widths for target: (a), (b) Gaussian spectra; (c), (d) triangular spectra; (e), (f) rectangular spectra. The pump field intensities after the final experimental optimization have been shown as insets.}
\label{fig:exp_result_l5mm}
\end{figure*}
\begin{table*}[htbp]
\centering
\caption{ \label{tab:l5mmcoeff} Optimized superposition coefficients $(\alpha_p)$ for  $L=5$ mm}
\begin{ruledtabular}
\begin{tabular}{c c ccccc}

\multirow{2}{*}{Spectrum}       & \multirow{2}{*}{Width} & \multicolumn{5}{c}{Complex coefficients}                                                                                                                                          \\ 
                             &                       & \multicolumn{1}{c}{$\alpha_0$}      & \multicolumn{1}{c}{$\alpha_1$}      & \multicolumn{1}{c}{$\alpha_2$}      & \multicolumn{1}{c}{$\alpha_3$}      & $\alpha_4$     \\ \hline
\multirow{2}{*}{Gaussian}    & 20                & \multicolumn{1}{c}{$-0.68 - 0.68i$} & \multicolumn{1}{c}{$0.14 + 0.01i$}  & \multicolumn{1}{c}{$0.00 + 0.00i$}  & \multicolumn{1}{c}{$0.24 + 0.00i$}  & $0.00 + 0.00i$ \\ 
                             & 30                & \multicolumn{1}{c}{$-0.56 - 0.56i$} & \multicolumn{1}{c}{$0.31 - 0.31i$}  & \multicolumn{1}{c}{$0.00 + 0.00i$}  & \multicolumn{1}{c}{$0.44 + 0.00i$}  & $0.01 + 0.00i$ \\ \hline
\multirow{2}{*}{triangular}  & 100                   & \multicolumn{1}{c}{$-0.93-0.09i$}   & \multicolumn{1}{c}{$-0.13+0.00i$}   & \multicolumn{1}{c}{$0.22 - 0.13i$}  & \multicolumn{1}{c}{$-0.20 + 0.01i$} & $0.00 + 0.00i$ \\  
                             & 200                   & \multicolumn{1}{c}{$0.04+0.34i$}    & \multicolumn{1}{c}{$-0.55-0.04i$}   & \multicolumn{1}{c}{$0.42-0.13i$}    & \multicolumn{1}{c}{$-0.04 - 0.46i$} & $0.00 + 0.42i$ \\ \hline
\multirow{2}{*}{rectangular} & 100                   & \multicolumn{1}{c}{$0.23 + 0.11i$}  & \multicolumn{1}{c}{$-0.56 - 0.09i$} & \multicolumn{1}{c}{$-0.23 - 0.05i$} & \multicolumn{1}{c}{$0.18 - 0.07i$}  & $0.25 - 0.68i$ \\  
                             & 150                   & \multicolumn{1}{c}{$0.14 + 0.00i$}  & \multicolumn{1}{c}{$-0.39 + 0.04i$} & \multicolumn{1}{c}{$0.71 - 0.32i$}  & \multicolumn{1}{c}{$-0.28 + 0.21i$} & $0.14 + 0.28i$ \\                             
\end{tabular}
\end{ruledtabular}
\end{table*}

\section{Experimental results with $L=5$ mm }\label{supple:l5mm_exp}

In this section, we report our experimental results with  $L=5$ mm,  $\theta_p = 28.69^{\circ}$, and $N=5$. Figures~\ref{fig:exp_result_l5mm}(a) and \ref{fig:exp_result_l5mm}(b) show the target and experimentally observed Gaussian OAM Schmidt spectra with standard deviations $20$ and $30$. Figures~\ref{fig:exp_result_l5mm}(c) and \ref{fig:exp_result_l5mm}(d) are the OAM Schmidt spectra of triangular shapes with base widths $100$ and $200$. The rectangular spectrum of widths $100$ and $150$ are reported in Figs.~\ref{fig:exp_result_l5mm}(e) and \ref{fig:exp_result_l5mm}(f), respectively. The intensity profiles of the pump field for each spectrum are shown as insets, and the corresponding $\alpha_p$ values are presented in Table~\ref{tab:l5mmcoeff}. Experimentally obtained $K_a$,  $E_f$, and  $G$ for each generated OAM Schmidt spectrum are reported in the respective figures.

\begin{table}[]
\centering
\caption{\label{tab:l0mmcoeff_8mode} Optimized superposition coefficients $(\alpha_p)$ with $L=10 mm$ and $N=8$ reported in Fig.~\ref{L10mm_spectrumandpump}(d) of the manuscript.}
\begin{ruledtabular}
\begin{tabular}{cc}
\multicolumn{2}{c}{Complex coefficients} \\ \hline
\multicolumn{1}{c}{\qquad $\alpha_0$}  &  $-0.08+0.68i$  \\ 
\multicolumn{1}{c}{\qquad $\alpha_1$}        &  $-0.01 -0.37i$ \\ 
\multicolumn{1}{c}{\qquad $\alpha_2$}        & $0.33-0.36i $\\ 
\multicolumn{1}{c}{\qquad $\alpha_3$}        & $-0.01-0.20i$ \\ 
\multicolumn{1}{c}{\qquad $\alpha_4$}        & $-0.08-0.09i $      \\ 
\multicolumn{1}{c}{\qquad $\alpha_5$}        &    $0.18 +0.00i $ \\ 
\multicolumn{1}{c}{\qquad $\alpha_6$}        &          $-0.19 + 0.12i $      \\ 
\multicolumn{1}{c}{\qquad $\alpha_7$}        &        $-0.10 +0.09i $     \\ 
\end{tabular}
\end{ruledtabular}
\end{table}
\begin{figure}[t!]
\centering
\includegraphics[scale=0.83]{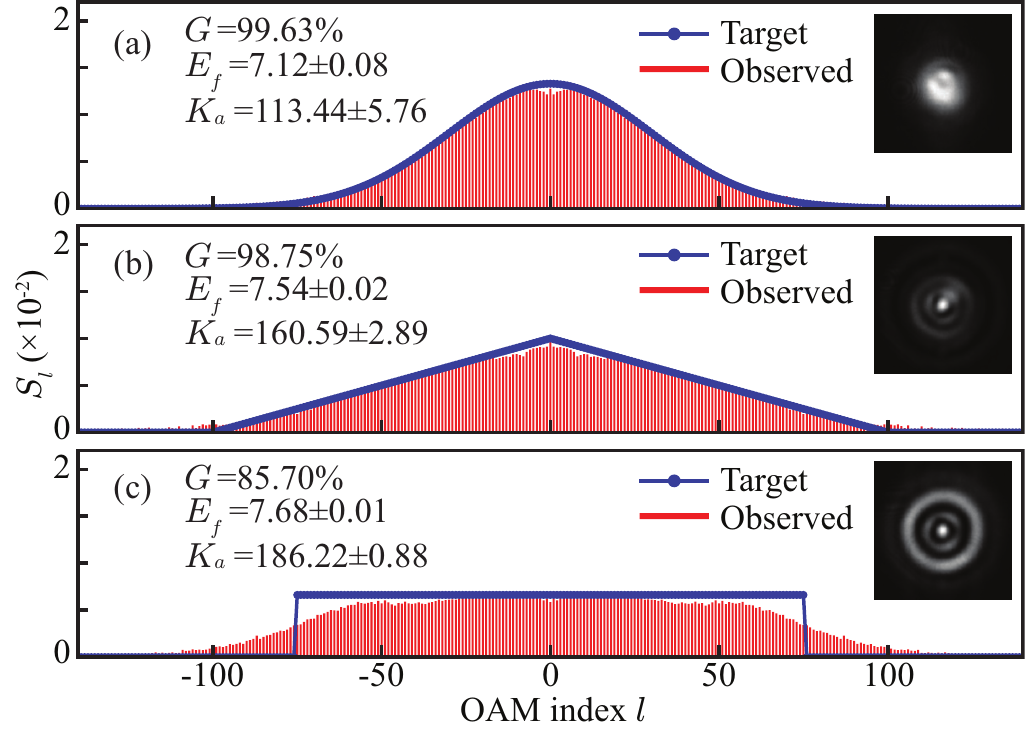}
\caption{Target and experimentally observed OAM Schmidt spectra with $L =10$ mm and $\theta_p= 28.71^{\circ}$ for the target (a) Gaussian spectrum, (b) triangular spectum, and (c) rectangular spectrum. The pump field intensities after the final experimental optimization have been shown as insets.}
\label{fig:exp_result_l10mm}
\end{figure}
\begin{table*}[]
\centering
\caption{ Optimized superposition coefficients $(\alpha_p)$ for $L=10$ mm.}
\begin{ruledtabular}
\begin{tabular}{ccccccc}
\multirow{2}{*}{Spectrum}       & \multirow{2}{*}{Width} & \multicolumn{5}{c}{Complex coefficients}                                                                                                                                          \\ 
                             &                       & \multicolumn{1}{c}{$\alpha_0$}      & \multicolumn{1}{c}{$\alpha_1$}      & \multicolumn{1}{c}{$\alpha_2$}      & \multicolumn{1}{c}{$\alpha_3$}      & $\alpha_4$     \\ \hline
\multirow{2}{*}{Gaussian}    & 20                & \multicolumn{1}{c}{$0.21+0.94i$} & \multicolumn{1}{c}{$-0.13+0.19i$}  & \multicolumn{1}{c}{$-0.09-0.01i$}  & \multicolumn{1}{c}{$-0.01-0.03i$}  & $0.03-0.02i$ \\ 
                             & 30                & \multicolumn{1}{c}{$-0.46-0.71i$} & \multicolumn{1}{c}{$0.43-0.22i$}  & \multicolumn{1}{c}{$0.06+0.20i$}  & \multicolumn{1}{c}{$-0.04+0.06i$}  & $-0.05-0.03i$ \\ \hline
\multirow{2}{*}{triangular}  & 100                   & \multicolumn{1}{c}{$0.55-0.57i$}   & \multicolumn{1}{c}{$0.47-0.31i$}   & \multicolumn{1}{c}{$-0.12-0.12i$}  & \multicolumn{1}{c}{$-0.14-0.05i$} & $-0.07-0.02i$ \\  
                             & 200                   & \multicolumn{1}{c}{$0.32+0.36i$}    & \multicolumn{1}{c}{$-0.10-0.68i$}   & \multicolumn{1}{c}{$-0.10+0.04i$}    & \multicolumn{1}{c}{$-0.04-0.01i$} & $-0.51-0.15i$ \\ \hline
\multirow{2}{*}{rectangular} & 100                   & \multicolumn{1}{c}{$-0.69+0.15i$}  & \multicolumn{1}{c}{$0.39-0.10i$} & \multicolumn{1}{c}{$0.39-0.34i$} & \multicolumn{1}{c}{$0.22-0.09i$}  & $-0.01+0.11i$ \\  
                             & 150                   & \multicolumn{1}{c}{$0.11+0.18i$}  & \multicolumn{1}{c}{$-0.44-0.59i$} & \multicolumn{1}{c}{$0.04+0.26i$}  & \multicolumn{1}{c}{$0.48+0.34i$} & $-0.02-0.02i$ \\ 
\end{tabular}
\label{tab:l0mmcoeff}
\end{ruledtabular}
\end{table*}

\begin{figure*}[]
\centering
\includegraphics[scale=0.86]{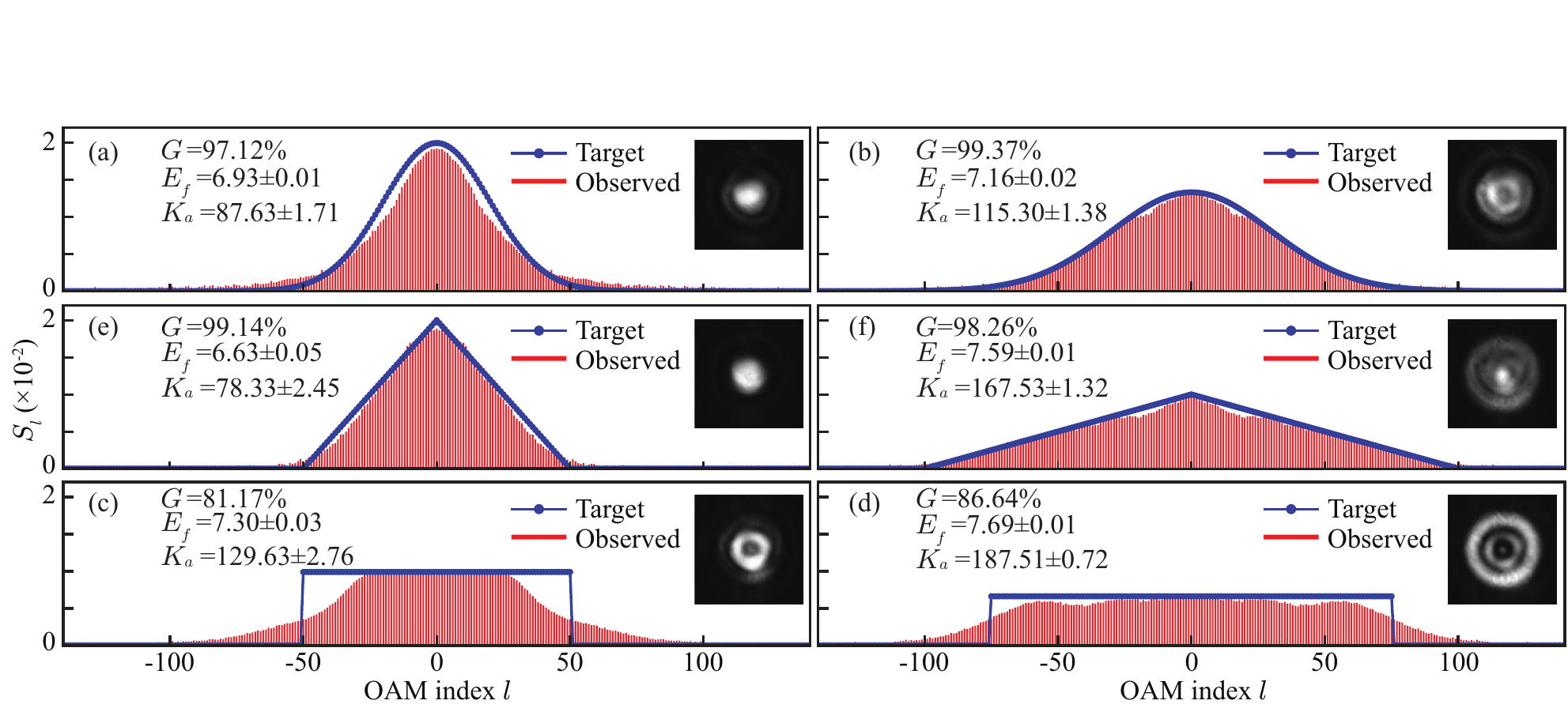}
\caption{ Target and experimentally observed OAM Schmidt spectra with $L =15$ mm, and $\theta_p= 28.71^{\circ}$ with two separate widths for target: (a), (b) Gaussian spectra; (c), (d) triangular spectra; (e), (f) rectangular spectra. The pump field intensities after the final experimental optimization have been shown as insets.}
\label{fig:exp_result_l15mm}
\end{figure*}
\begin{table*}[]
\centering
\caption{ \label{tab:l15mmcoeff} Optimized superposition coefficients $(\alpha_p)$ for $L=15$ mm}
\begin{ruledtabular}
\begin{tabular}{ccccccc}

\multirow{2}{*}{Spectrum}       & \multirow{2}{*}{Width} & \multicolumn{5}{c}{Complex coefficients}                                                                                                                                           \\  
                             &                       & \multicolumn{1}{c}{$\alpha_0$}      & \multicolumn{1}{c}{$\alpha_1$}      & \multicolumn{1}{c}{$\alpha_2$}      & \multicolumn{1}{c}{$\alpha_3$}      & $\alpha_4$      \\ \hline
\multirow{2}{*}{Gaussian}    &  20                & \multicolumn{1}{c}{$-0.17+0.82i$}   & \multicolumn{1}{c}{$0.34-0.03i$}    & \multicolumn{1}{c}{$0.34-0.03i$}    & \multicolumn{1}{c}{$-0.17 + 0.00i$} & $-0.20 + 0.03i$ \\  
                             &  30                & \multicolumn{1}{c}{$-0.78 + 0.43i$} & \multicolumn{1}{c}{$0.31 - 0.03i$}  & \multicolumn{1}{c}{$0.25 - 0.03i$}  & \multicolumn{1}{c}{$-0.17 + 0.00i$} & $-0.14 + 0.03i$ \\ \hline
\multirow{2}{*}{triangular}  & 100                   & \multicolumn{1}{c}{$0.19 + 0.97i$}  & \multicolumn{1}{c}{$0.02+ 0.01i$}   & \multicolumn{1}{c}{$-0.09-0.09i$}   & \multicolumn{1}{c}{$0.06+ 0.06i$}   & $-0.01 + 0.02i$ \\  
                             & 200                   & \multicolumn{1}{c}{$0.03 + 0.32i$}  & \multicolumn{1}{c}{$-0.87 - 0.03i$} & \multicolumn{1}{c}{$0.03 - 0.15i$}  & \multicolumn{1}{c}{$-0.15 - 0.26i$} & $0.00 + 0.15i$  \\ \hline
\multirow{2}{*}{rectangular} & 100                   & \multicolumn{1}{c}{$0.83 + 0.00i$}  & \multicolumn{1}{c}{$0.31 + 0.18i$}  & \multicolumn{1}{c}{$-0.14 + 0.13i$} & \multicolumn{1}{c}{$0.00 + 0.35i$}  & $0.00 + 0.18i$  \\  
                             & 150                   & \multicolumn{1}{c}{$0.20+0.20i$}    & \multicolumn{1}{c}{$-0.83+0.00i$}   & \multicolumn{1}{c}{$0.22-0.23i$}    & \multicolumn{1}{c}{$0.10 - 0.33i$}  & $-0.10 - 0.02i$ \\ 
\end{tabular}
\end{ruledtabular}
\end{table*}

\section{Experimental results with $L=10$ mm }\label{supple:l10mm_exp}

In this section, we report our experimental results with  $L=10$ mm, $\theta_p = 28.71^{\circ}$, and $N=5$. Figure~\ref{fig:exp_result_l10mm}(a), Figure~\ref{fig:exp_result_l10mm}(b), Figure~\ref{fig:exp_result_l10mm}(c) show the plot of target and experimentally observed $S_l$ for a Gaussian spectrum with standard deviation $30$, a triangular spectrum with base width $200$, and a rectangular spectrum of width $150$. The spatial intensity profiles of the pump field for each spectrum are shown as insets, and the corresponding $\alpha_p$ values are presented in Table~\ref{tab:l0mmcoeff}, which also provides the optimized $\alpha_p$ values for a Gaussian spectrum of standard deviation $20$, a triangular spectrum of base width $100$, and a rectangular spectrum with width $100$. In Figure~\ref{L10mm_spectrumandpump}(d) of the manuscript, we reported our observations with  $L=10$ mm and $N=8$. The corresponding superposition coefficients $\alpha_p$ are reported in Table~\ref{tab:l0mmcoeff_8mode}.

\section{Experimental results with $L=15$ mm }\label{supple:l15mm_exp}
In this section, we report our experimental results with $L=15$, $\theta_p = 28.71^{\circ}$ and $N=5$. Figures~\ref{fig:exp_result_l15mm}(a) and \ref{fig:exp_result_l15mm}(b) show the target and experimentally observed Gaussian OAM Schmidt spectra with standard deviations $20$ and $30$. Figures~\ref{fig:exp_result_l15mm}(c) and \ref{fig:exp_result_l15mm}(d) are the OAM Schmidt spectra of triangular shapes with base widths $100$ and $200$. The rectangular spectrum of widths $100$ and $150$ are reported in Figs.~\ref{fig:exp_result_l15mm}(e) and \ref{fig:exp_result_l15mm}(f), respectively. The spatial intensity profiles of the pump field are shown as insets, and the corresponding $\alpha_p$ values are presented in Table~\ref{tab:l15mmcoeff}.

\bibliography{postselection-free_ref}

\end{document}